\documentclass[%
 reprint,
superscriptaddress,
 amsmath,amssymb,
 aps,
pra,
]{revtex4-2}

\usepackage{graphicx}
\usepackage{subcaption}
\usepackage{dcolumn}
\usepackage{bm}
\usepackage[colorlinks=true,linkcolor=blue,citecolor=blue]{hyperref}

\usepackage{amsthm}
\usepackage{amsmath}
\usepackage{braket}
\usepackage{dsfont}
\usepackage[dvipsnames,table,xcdraw]{xcolor}
\usepackage{tikz-cd}
\usetikzlibrary{positioning}

\newtheorem{theorem}{Theorem}

\newtheorem{coro}{Corollary}

\newcommand{\chn}{\varphi} 
\newcommand{\rf}{\gamma} 

\newcommand{\pp}{\chn p}

\newcommand{\mt}[1]{M^{#1}}
\renewcommand{\vr}[1]{v^{#1}}
\newcommand{\rvp}{\mathcal{R}_{\chn,\rf}}
\newcommand{\ret}[2]{\mathcal{R}_{#1,#2}}

\newcommand{\cevi}{q} 

\hypersetup{
    colorlinks=true,
    linkcolor=blue,
    filecolor=magenta,      
    urlcolor=blue,
    pdftitle={Overleaf Example},
    pdfpagemode=FullScreen,
    }

\begin{document}

\title{Rational v. Radical: Trajectory Properties of Iterated Bayesian Updates}
\author{Clive Cenxin Aw}%
\email{clive.awcx@ntu.edu.sg}
\affiliation{School of Physical and Mathematical Sciences, Nanyang Technological University, 637371, Singapore}
\affiliation{Centre for Quantum Technologies, National University of Singapore, 3 Science Drive 2, 117543, Singapore}

\author{Lynette Zhong Ling Wang}%
\email{wzllynette@gmail.com} 
\affiliation{Singapore Management University, 81 Victoria St, Singapore 188065}

\author{Peter Sidajaya}%
\email{ptrsdjy@gmail.com}
\affiliation{Centre for Quantum Technologies, National University of Singapore, 3 Science Drive 2, 117543, Singapore}


\date{\today}

\begin{abstract}
Iterating Jeffrey's update on fixed evidence gives a maximum-likelihood estimate under suitable support conditions. What is less clear would be how the trajectories are organized, and what happens when we trace them in reverse. For a fixed strictly positive, invertible classical channel and full-support evidence, reverse iteration corresponds to repeated prior hacking. Almost every interior prior then approaches certainty in one possibility while each nontrivial step worsens agreement with the evidence. We call this limiting behavior radicalization. Priors that approach the same pure state form a Bayes phase. In the phase diagrams, the separating trajectories instead approach facet fixed points determined by M-projections. We also ask whether one can choose evidence that brings every interior prior to the same conclusion. For a strictly positive, invertible channel this is always possible, although one cannot always prescribe which conclusion it will be.
\end{abstract}

\maketitle

\section{Introduction}

It is common knowledge that iterating Bayesian updates on some given evidence asymptotically recovers the conclusions of frequentist methods. In the classical setting considered here, repeated updating against fixed evidence is an expectation--maximization procedure whose limit, under suitable support conditions, is a maximum-likelihood estimate \cite{lucy1974rectification,shepp1982maximum,pinzon2025jeffrey, murk2026connectingquantumtomographyquantum, sidajaya2026tomographic-negative-paper}. This tells us only where the updates lead, as opposed to the journey they take. That is, not much has been said about the \textit{trajectories} of such Bayesian updates, how they are organized across the space of priors, or what happens when we trace them backward. These become particularly interesting when we envisage the distributions as beliefs and the updates as rational responses to encountered evidence.

In relation to these notions, closed-mindedness has been of recent interest in epistemology \cite{Battaly2020-dogma-philo,cassam2016vice-dogma-philo,battaly2018dogmatism,battaly2018virtue,dunne2025vice} and sociology \cite{savion2009clinging-dogma-social,siebert2023effective-dogma-social,nguyen2020echo,taber2006motivated}. Such behaviors may be modeled by an agent who refuses to change his prior belief on some matter, even when its evidential basis has been discredited. Easily, this kind of dogmatism is antithetical to Bayesian reasoning, with Bayes' rule as centerpiece, which is widely regarded as \textit{the} prescription for rational responsiveness to counter-evidence \cite{jaynes2003,jeffrey1990logic-BK,chan2005revision-BK,jacobs-changing-mind, chivers2024-bayes-app}. Famously, it formalizes how an agent should revise their prior information given observed evidence and the conditional probabilities (which relate that evidence to the possibilities under consideration). This framework is part of standard probability theory \cite{feller1968introduction} and finds applications across the sciences and machine learning \cite{jessop2018-bayes-app,chivers2024-bayes-app,tipping2003bayesian-bayes-app}.

It turns out that, nevertheless, Bayes' rule can be invoked \textit{while being closed-minded}, albeit in a conceptually inverted way. As shown with greater detail in \cite{aw-sidajaya-hacking}, one can (generically) begin with a desired conclusion and engineer a reference prior that produces it under Bayesian updating. This is \emph{prior hacking}. When the channel and evidence are given, the claimed prior is tuned so as to acquire a preselected ``update''. Thus, an agent can appear to update rationally while really protecting a conclusion chosen in advance \footnote{This feature of Bayes' rule can also be connected to another rational update known as a Schr{\"o}dinger Bridge \cite{aw-sidajaya-hacking,friedland2017schrodinger,orland2025schrodinger-SB-gen}}. 

\begin{figure*}[t]
    \centering
    \includegraphics[width=\textwidth]{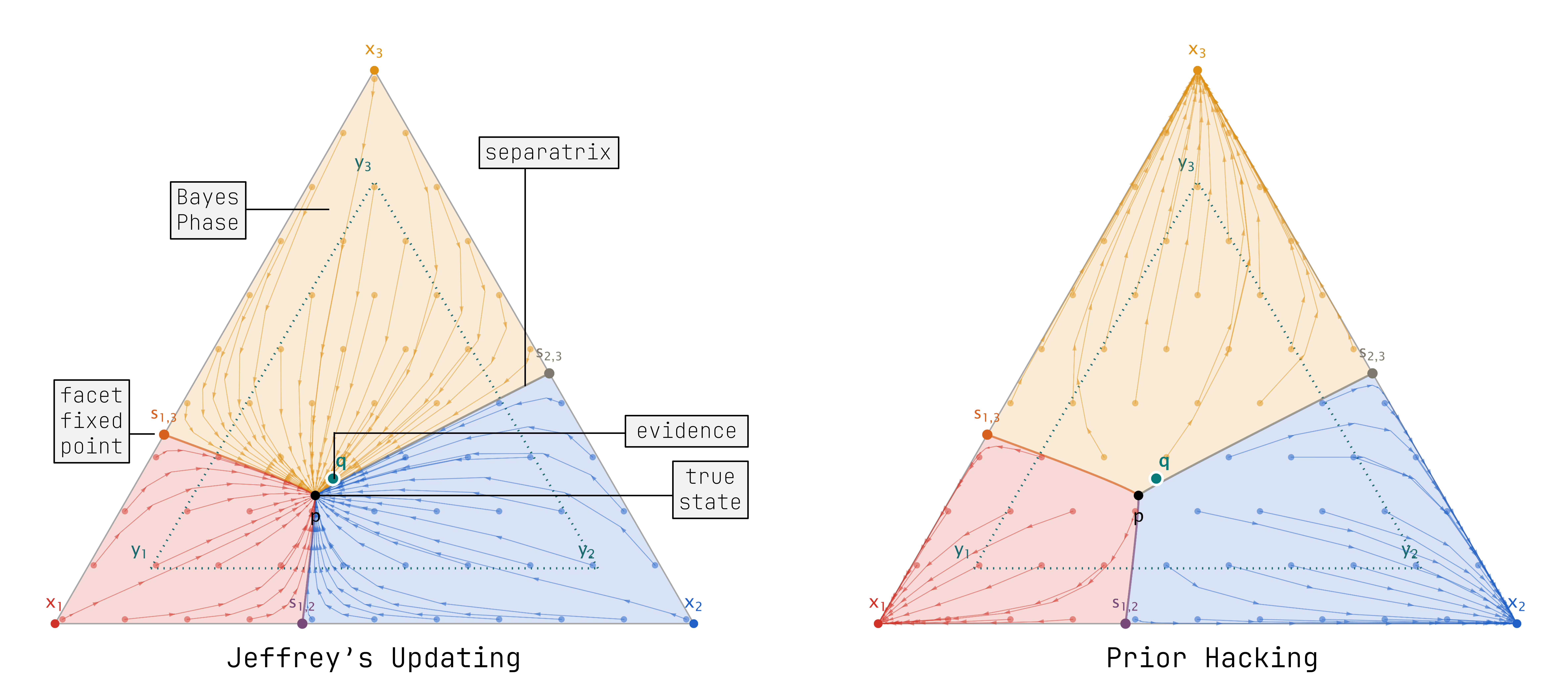}
    \caption{Iterated Jeffrey updating (left) and prior hacking (right) on the trit simplex $\Delta$, for the same fixed channel $\chn$ (with image $\chn(\Delta)$ depicted as the cyan dotted triangle with vertices $y_a=\chn\delta_{x_a}$) and evidence $q=\chn p$. Vertices labeled $x_a$ represent the pure states $\delta_{x_a}$; arrows indicate successive iterations. Jeffrey updating sends interior priors toward $p$ (Theorem~\ref{thm:cl-emb-inv}), whereas prior hacking reverses these trajectories, generically approaching pure states via radicalization (Corollary~\ref{coro:almost-all-radicalize}). The shaded regions are the Bayes phases $\mathcal{B}_{x_a}$ (Sec.~\ref{sec:bayes-phases}), colored by the pure state approached under iterated prior hacking. The separatrices $\mathcal{P}_S$ (Sec.~\ref{sec:swing-party}, Eq.~\eqref{eq:party-line}) are the separating curves. Priors on them approach the corresponding non-pure facet fixed points $s_S$ (Eq.~\eqref{eq:swing-mprojection}), labeled $s_{a,b}$ on the $x_a$--$x_b$ edges. The state $p$ remains fixed in both directions (Theorem~\ref{thm:radic-or-party}), and the same phase coloring is used in both panels. Representative instances of these categories are labeled for ease of reference; not every instance is labeled, to avoid encumbering the figure (this also applies to Figures \ref{fig:trit23} and \ref{fig:quart1}).}
    \label{fig:trit1}
\end{figure*}

Now, since such hacking is generically possible \cite{aw-sidajaya-hacking}, a further question arises: what happens when this is done repeatedly? The prior obtained in one round of hacking can itself become the new conclusion to be justified in the next. Doing so traces the corresponding Bayesian trajectory \textit{backward}. How, then, can agents facing the same evidence end up certain of different things?

Following recent work connecting tomography and retrodiction \cite{murk2026connectingquantumtomographyquantum,sidajaya2026tomographic-negative-paper}, we fix a classical channel $\chn$ and repeatedly apply Jeffrey's update to the same evidence distribution $q$. Only the reference prior changes between iterations. This provides a concrete setting within the broader study of belief kinematics \cite{jeffrey1990logic-BK,zhou2014belief-BK,spohn1988ordinal-BK,van1980rational-BK,bradley2007kinematics-BK}. Analogically speaking, as much as frictionless planes isolate features of mechanical motion, fixing the channel and evidence isolates the consequences of revising, or engineering, one's prior. The emergent features, while certainly not corresponding one-to-one with social phenomena, provide illustrative formalized counterparts to them.

Notably, this work shows that for strictly positive, invertible channels and evidence $q$ (which may or may not be generated by an interior state $p$, such that $q$ is $p$ propagated through $\chn$, written here as $q =\chn p$), the simplex interior decomposes into \emph{Bayes phases},  \emph{separatrices}, and (if $q$ is in the image of $\chn$) the input $p$. Each Bayes phase contains the priors that approach the same pure state under iterated prior hacking. We call this \emph{radicalization}: the limiting belief assigns certainty to one possibility. Priors on the separatrices instead approach facet fixed points. These exceptional sets have measure zero, so almost every interior prior radicalizes under repeated prior hacking. Thus, while forward updating recovers the evidence-compatible state, its systematic reversal almost always leads to certainty in a single alternative. These sociologically evocative behaviors arise from the geometry of the updating rule itself.

This work is sectioned as follows. Section~\ref{sec:jeffrey} considers the forward direction of iterated Jeffrey updating, including recovery of the true state and the stepwise improvement in agreement with the evidence. Section~\ref{sec:hacking} then traces these trajectories backward through prior hacking. We classify their possible limits (Theorem~\ref{thm:radic-or-party}), introduce how the notion of Bayes phases naturally emerges from this, and show that almost every interior prior radicalizes (Theorem~\ref{thm:party-measure-zero} and Corollary~\ref{coro:almost-all-radicalize}). We also consider whether shared evidence can bring every interior prior to the same conclusion, and the limits on prescribing that conclusion. Section~\ref{sec:discussion} discusses these results, before we conclude in Section~\ref{sec:conclusion}. Full proofs are typically collected in Appendix~\ref{app:proofs}, except where given directly or referred to existing work. For added understanding, proof sketches are sometimes included in the main text, especially for more involved arguments. 

\begin{figure*}[t]
    \centering
    \includegraphics[width=1.\linewidth]{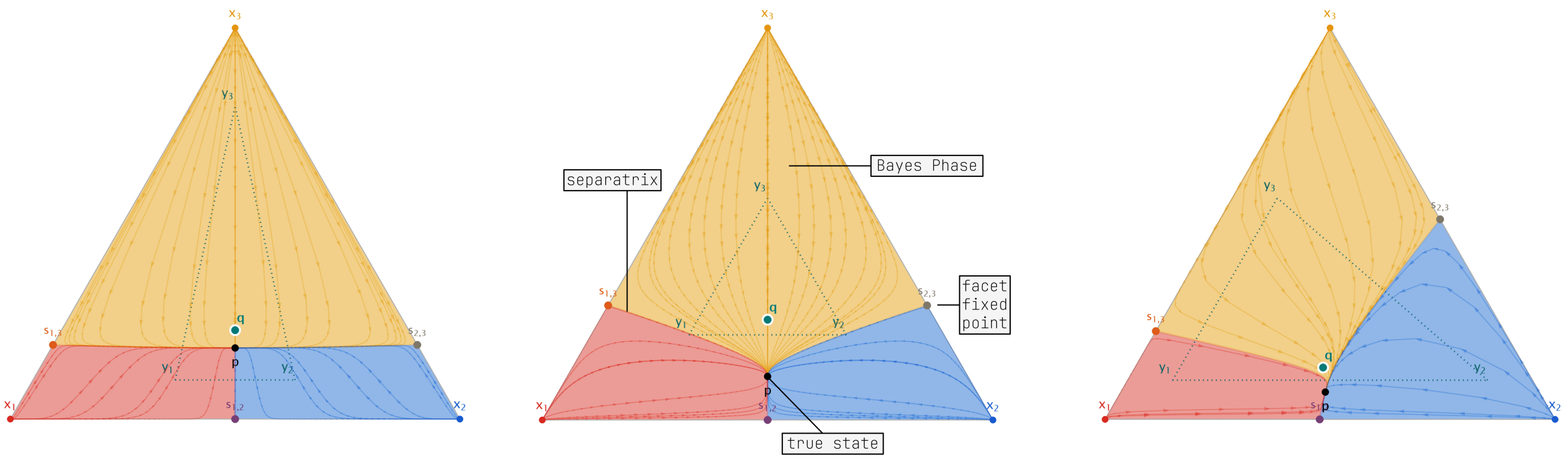}
    \caption{Same conventions as Figure \ref{fig:trit1}, but with three other choices of $\chn$. We plot only the Jeffrey updating direction (indicated by the arrowheads), as prior hacking goes in the opposite direction. Bayes phases and their separatrices are emphasized here with higher opacity as well. Notice how the trajectories tend to advance at a slower pace where the channel has reduced distinguishability.}
    \label{fig:trit23}
\end{figure*}

\section{Trajectory Properties of Jeffrey's Update} \label{sec:jeffrey}
Before considering prior hacking, we consider the trajectory properties of iterated Jeffrey's updating (which has been of recent interest in the context of state tomography \cite{sidajaya2026tomographic-negative-paper,murk2026connectingquantumtomographyquantum}). We begin by formalizing the problem. Classical transformations or channels are represented by stochastic maps $\chn$ defined by $d' \times d$ transitions $\chn(y|x)$ that act on distributions $p(x)$, which represent descriptions of classical inputs. These produce output distributions
\begin{equation}
    [\chn p](y)=\sum_x \chn(y|x)p(x). 
\end{equation}
Consequently, the input space is defined by $x \in \{1,...,d\}$ and the output space by $y \in \{1,...,d'\}$. We restrict to square channels in view of invertibility requirements. 

 Now, Bayes' rule enforces the insufficiency of $\chn$ for making an inference about the input given some evidence $\cevi$. One needs not only forward transitions given by $\chn$ but also some reference distribution $\rf$ \cite{AwBS, watanabe65}. All this to say that under Bayes' rule, given $\cevi$, we update our prior information (encoded in $\rf$) to a new description
 $\rvp \cevi$, where $\rvp$ is the map corresponding to retrodiction: 
 \begin{equation} \label{bayes}
    \rvp(x|y)=\chn(y|x)\frac{\rf(x)}{[\chn\rf](y)}.
\end{equation}
A Bayesian update from a prior $\rf_i$ to a new conclusion $\rf_{i+1}$ is given by what is called Jeffrey's conditionalization. Observing output evidence $q$, we send it through the Bayesian inverse as follows. 
\begin{equation}\label{eq:jeff}
    \ret{\chn}{\rf_i}q=\rf_{i+1}
\end{equation}

It is well known that, under iteration of Eq.~\eqref{eq:jeff} with fixed $\chn,q$, $\rf_i$ eventually converges to the maximum-likelihood estimator with respect to the data $q$. In this case, it is more accurate to write $\rf_i^{\chn,q}$ for \eqref{eq:jeff}, for all $i\in \mathbb{Z}^{>0}$, as $\rf_1^{\chn,q}$ can be induced by some arbitrarily chosen $\rf_0$. To avoid encumbering the notation, however, we will not include these superscripts unless the context is insufficiently clear.

Of note, when the channel is injective and the evidence is within the image of the channel, such an update recovers the ``true'' state. Formally, 

\begin{theorem} \label{thm:cl-emb-inv}
     For a channel $\chn$, invertibility is equivalent to recovery of every input distribution $p$ from its evidence $\pp$, for every initial prior whose support contains that of $p$. Formally,
     \begin{equation}
       \chn^{-1}\text{ exists}\quad\Longleftrightarrow\quad
       \begin{gathered}
         \forall(p,\rf_0)\text{ s.t.} \operatorname{supp}p\subseteq\operatorname{supp}\rf_0:\\
         \lim_{i\to\infty}\rf_i=p
       \end{gathered}
     \end{equation}
\end{theorem}
\begin{proof}
    A direct proof can be found in \cite{sidajaya2026tomographic-negative-paper}. 
    \newline
\end{proof}
 \subsection{Stepwise Features of Iterated Jeffrey's Update}

In addition to this property, other stepwise features of Jeffrey's update may be noted. Importantly,

\begin{theorem}\label{thm:cpop-ineq}
For fixed $\chn$ and evidence $q$, write
\begin{equation}\label{eq:divergence}
    \mathcal{D}(\rf):=D_{\mathrm{KL}}(q\Vert\chn\rf)
    =\sum_y q(y)\log\frac{q(y)}{[\chn\rf](y)}.
\end{equation}
This measures the divergence from the evidence to the output predicted by the prior $\rf$.
The following post-processing inequality holds:
\begin{equation}\label{eq:cpop-ineq}
    \forall(\chn,p,\rf_0, i\in \mathbb{Z}_{\geq0}): \mathcal{D}(\rf_{i+1}) \leq \mathcal{D}(\rf_{i})
\end{equation}
with equality only if $\rf_{i+1}=\rf_{i}.$
\end{theorem}
\begin{proof}
   See Appendix \ref{app:cpop-ineq}
\end{proof}

This implies that every inference on repeated evidence results, at worst, in no progress. Otherwise, it results in a better fit between it and one's prior. Now, we consider some other results when $q=\pp$ specifically \footnote{These results cannot be generalized to arbitrary $q$ because $q$ can be totally decoupled from $p$ (e.g., a poor sample of $\pp$).}. First, we consider a corresponding pre-processing inequality:
\begin{theorem}\label{thm:cl-prp-ineq}
    When $q=\pp$, each update never brings the prior further from $p$ in KL divergence:
    \begin{equation}\label{prp-ineq}
        \forall(\chn,p,\rf_0, i\in \mathbb{Z}_{\geq0}): {D_{\mathrm{KL}}}\big(p\Vert \rf_{i+1}\big)\le{D_{\mathrm{KL}}}\big(p\Vert \rf_i\big),
    \end{equation}
\end{theorem}
\begin{proof}
 
From \cite{li-winter}, we have the following theorem:
\begin{equation}
   {D_{\mathrm{KL}}}\big(p\Vert \rf_i\big) - D_{\mathrm{KL}}(\pp \Vert  \chn\rf_i)\ge   {D_{\mathrm{KL}}}\big(p\Vert \rf_{i+1}\big)
\end{equation}
\eqref{prp-ineq} follows immediately, since $ \forall(p,q):{D_{\mathrm{KL}}}(p\Vert q) \geq 0$. 
\end{proof}

Furthermore, this also implies that as long as $\pp \neq \chn\rf_i$, \eqref{prp-ineq} becomes a strict inequality.
\begin{coro} \label{coro:cl-strict-progress} As long as the propagation of our current prior $\chn \rf_i$ is not the same as the evidence $\pp$, the KL divergence from $p$ to the prior strictly decreases.
    \begin{equation} \label{eq:cl-strict-progress}
    \pp \neq \chn\rf_i \quad \Rightarrow \quad {D_{\mathrm{KL}}}\big(p\Vert \rf_{i+1}\big) <  {D_{\mathrm{KL}}}\big(p\Vert \rf_i\big) 
\end{equation}
\end{coro}
From this, another corollary emerges. Denoting a column-stochastic matrix representation of $\chn$ as $\mt{\chn}$ and stochastic vectors representing $p$ as $\vr{p}$, we have:
\begin{coro} \label{coro-kern}
For $i\in \mathbb{Z}_{>0}$, when there are no longer improvements in iterating Bayesian updates on some (repeatedly received) evidence $\pp$, then the retrodiction protocol's effect is in the kernel of $\chn$:
    \begin{align}
        &{D_{\mathrm{KL}}}\big(p\Vert \rf_{i+1}\big) =  {D_{\mathrm{KL}}}\big(p\Vert \rf_i\big) \nonumber \\  &\Rightarrow \mt{\ret{\chn}{\rf_{i-1}}}\mt\chn \vr p= \vr p + \vr \delta,  \label{eq:coro-kern}  \\ & \quad\text{where } \vr \delta \in \ker(\mt\chn).\nonumber 
    \end{align}        
\end{coro}
\begin{proof}
From the contrapositive of Corollary \ref{coro:cl-strict-progress},
\begin{align*}
    & {D_{\mathrm{KL}}}\big(p\Vert \rf_{i+1}\big) =  {D_{\mathrm{KL}}}\big(p\Vert \rf_{i}\big) \\  &\Rightarrow \; \pp = \chn\rf_i \\
    & \Leftrightarrow \; \mt \chn \vr p = \mt \chn \mt {\ret{\chn}{\rf_{i-1}}}\mt \chn \vr p \\
    & \Leftrightarrow \;  \mt \chn (\mathbb{I} - \mt {\ret{\chn}{\rf_{i-1}}}\mt \chn) \vr p = 0 \\
    & \Leftrightarrow \;  (\mathbb{I} - \mt {\ret{\chn}{\rf_{i-1}}}\mt \chn) \vr p \in \ker(\mt \chn)
\end{align*}
which is simply \eqref{eq:coro-kern}.
\end{proof}
To put Corollary \ref{coro-kern} more conceptually, we might say that when there is no longer a better fit with the evidence, it is because the update is being induced by the prior alone. That is, the process is playing no role (i.e. the update is in its kernel), and so Bayes' rule simply halts further updating. 

\begin{figure}[t]
    \centering
    \includegraphics[width=0.97\linewidth]{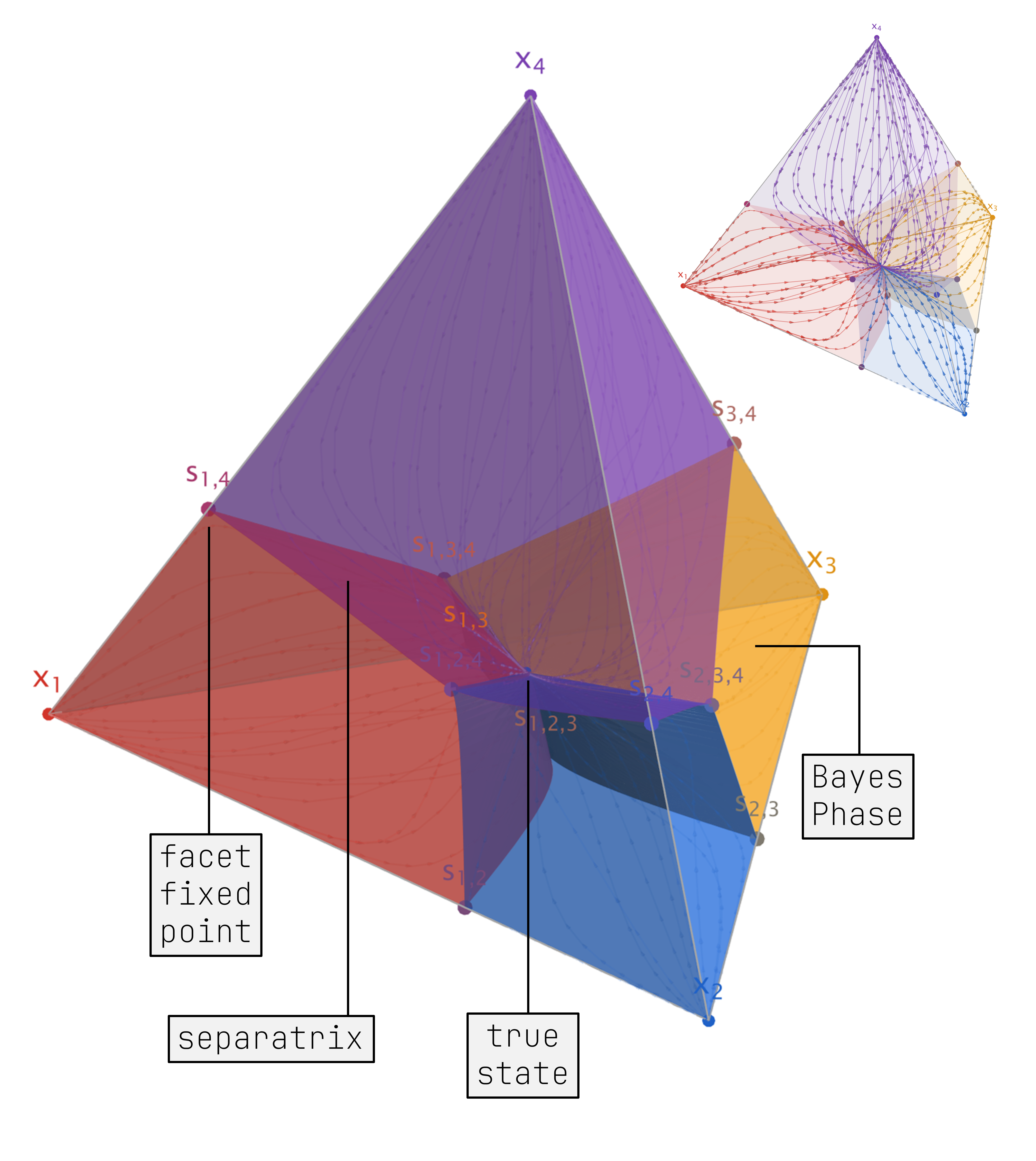}
    \caption{Same conventions as Figure~\ref{fig:trit23}, now for a generic quart channel. The upper-right inset highlights the Bayesian trajectories, while the larger diagram shows the arrangement of Bayes phases and their bounding separatrices. Notice how the separatrices here form two-dimensional surfaces.}
    \label{fig:quart1}
\end{figure}

\section{Trajectory Properties of Prior Hacking}\label{sec:hacking}
One can write Jeffrey's update \eqref{eq:jeff} as a map $F_{\chn,q}$ acting on $\rf_i$ based on the conditional and evidence $(\chn,q)$:
\begin{equation}
    F_{\chn,q}(\rf_i)=\ret{\chn}{\rf_i} q=\rf_{i+1},
\end{equation}
which in turn gives $F_{\chn,q}^j (\rf_i) = \rf_{i+j}$. When $\chn$ is invertible and $q=\pp$, Theorem~\ref{thm:cl-emb-inv} gives recovery of $p$ from any interior prior via $F_{\chn,q}^
\infty(\rf_i)=p$. We now consider the inverse of Jeffrey's updating of priors. This is called ``prior hacking'' and may be written as such:
\begin{equation}
    F_{\chn,q}^{-1} (\rf_{i}) = \rf_{i-1}.
\end{equation}
As discussed in greater detail in \cite{aw-sidajaya-hacking}, this is nothing but the operation of, first, fixing a \textit{conclusion} $\rf_{i}$ of some Jeffrey's update based on the conditionals given by $\chn$ and some evidence $q$, and then finding the reference prior $\rf_{i-1}$ that would have reached that conclusion through \eqref{eq:jeff}. 

This is to say that through prior hacking an agent appears to do a rational belief update, while actually bearing a pre-established conclusion from the start. The exact conditions under which prior hacking is possible are delineated in \cite{aw-sidajaya-hacking}. When it is possible, finding the hacking prior is a matter of a simple Sinkhorn iteration process. How this is done is as follows: for a hacking problem characterized by the triple $(\chn,q,\rf_i)$, the hacking prior is given by $\rf_{i-1}=\rf_{h}$, for which $\rf_h$ is the distribution obtained when the following iteration converges:
\begin{equation}
    \rf_{h} = \rf_i \oslash [\chn^{\text{T}}\rf_{h}] \; \leftrightharpoons \;\rf_{h} = q \oslash [\chn\rf_{h}]
\end{equation}
We note that channels for which $\forall p \in X, \forall y :[\pp](y)>0$ (i.e., all outputs have full support) can always be prior hacked for any evidence $q$ and any conclusion $\rf_{i}$. It is for these full-support channels that we establish our theorems. With that, we turn to one of our key results.
\subsection{Facet Fixed Points, Separatrices and Radicalization}
\label{sec:swing-party}

We begin with some definitions. Let $\chn>0$ be an invertible stochastic matrix with input space $X$ and output space $Y$. Write $d=|X|=|Y|$ and let $\Delta$ be the input simplex, of dimension $d-1$. We write $S$ for a positive-dimensional proper face of $\Delta$. For a trit, these are the three edges. Each is obtained by setting one of the three input probabilities to zero.

For each such face $S$, we seek the input $z\in S$ whose image $\chn z$ best fits the evidence $q$. Specifically, we minimize the divergence from $q$ over $\chn(S)$ \cite{csiszar1975geometry} and denote its input by $s_S$:
\begin{equation}
s_S
=
\arg\min_{z\in S}
\mathcal{D}(z).
\label{eq:swing-mprojection}
\end{equation}
Since $\chn$ is injective, each face has a unique minimizer $s_S$. When $s_S$ lies in the relative interior of $S$, it is fixed under Jeffrey's update. We shall call $s_S$ the \emph{facet fixed point} of $S$. Here the supporting face $S$ may have any positive dimension below that of $\Delta$. 

Further, we denote as $\mathcal{P}_S$ the set of interior priors that converge to $s_S$ under iterated prior hacking. These are the \emph{separatrices} in the phase diagrams (Fig.~\ref{fig:trit1},\ref{fig:trit23} and \ref{fig:quart1}), noting that in higher dimensions they generally are not one dimensional (Fig.~\ref{fig:quart1}).
\begin{equation}
\mathcal{P}_S
=
\left\{
\gamma\in\operatorname{int}\Delta:
F^{-i}_{\chn,q}(\rf)\longrightarrow s_S
\right\}.
\label{eq:party-line}
\end{equation}

Another notable behavior is when iterated hacking sends an initializing prior $\rf$ to a pure state. That is,
\begin{equation}\label{eq:rad}
    F^{-i}_{\chn,q}(\rf) \longrightarrow \delta_x,
\end{equation}
for some pure state $\delta_x$. We shall call this behavior \textit{radicalization}. That is, $\rf$ is radicalized to $x$. With these all defined, we state a key result:
\begin{theorem} \label{thm:radic-or-party}
For strictly positive, invertible $\chn$, full-support $q$, and any $\rf\in\operatorname{int}\Delta$, exactly one of the
following obtains under iterated prior hacking (i.e. $F^{-i}_{\chn,q}(\rf)$ for large $i$):
\begin{enumerate}
    \item $\rf$ is the ``true'' state $p$ and $q=\chn p$.
    \item $\rf$ is on a separatrix \eqref{eq:party-line}, and thus goes to the corresponding facet fixed point.
    \item $\rf$ is radicalized \eqref{eq:rad}.
\end{enumerate}
\end{theorem}
\begin{proof}
Very briefly, Theorem~\ref{thm:cpop-ineq} shows that prior hacking strictly increases $D_{\mathrm{KL}}(q\Vert\chn\rf)$ unless the prior is fixed. The divergence is bounded, and its equality condition forces every accumulation point to be fixed. The possible limits are the evidence-compatible interior state $p$ when it exists, a facet fixed point, and a pure state. See Appendix~\ref{app:radic-or-party} for the proof proper.
\end{proof}

\subsection{Bayes Phases}
\label{sec:bayes-phases}

With this, we can call the basin of a pure state $\delta_x$ under iterated prior hacking its \emph{Bayes phase}. That is,
\begin{equation}
\mathcal{B}_x
=
\left\{
\gamma\in\operatorname{int}\Delta:
F^{-i}_{\chn,q}(\rf)\longrightarrow\delta_x
\right\}.
\end{equation}

When $q=\chn p$ for an interior $p$, Theorem~\ref{thm:radic-or-party} gives the partition
\begin{equation}
\operatorname{int}\Delta
=
\{p\}
\sqcup
\bigcup_x\mathcal{B}_x
\sqcup
\bigcup_S\mathcal{P}_S,
\end{equation}
where the Bayes phases contain the radicalizing priors and the separatrices
contain precisely those non-equilibrium priors converging to
non-pure facet fixed points. 
\vspace{1em}

\begin{theorem}\label{thm:party-measure-zero}
Let $\chn$ be an invertible, strictly positive channel and let
$q$ have full support. Every separatrix $\mathcal{P}_S$ associated with
a non-pure facet fixed point $s_S$ has measure zero in $\Delta$. Equivalently,
\begin{equation}
    \mu\left(\bigcup_S\mathcal{P}_S\right)=0,
\end{equation}
where $\mu$ is the $(d-1)$-dimensional Lebesgue measure on $\Delta$.
\end{theorem}
\begin{proof}
    Briefly, prior hacking expands certain deviations from each non-pure facet fixed point, so convergence to it requires fine-tuning. The center-stable manifold theorem \cite{wen2016differentiable,smale1967differentiable} confines the permitted nearby priors to a lower-dimensional set. Every prior on the separatrix eventually enters this set, and tracing back finitely many steps preserves measure zero. A countable union over these step counts, followed by a finite union over facet fixed points, gives the result. See Appendix~\ref{proof:party-measure-zero} for the full proof.
\end{proof}

\begin{coro}\label{coro:almost-all-radicalize}
Under the conditions of Theorem~\ref{thm:radic-or-party}, almost every
initial prior in $\operatorname{int}\Delta$ radicalizes under iterated
prior hacking.
\end{coro}

\begin{proof}
By Theorem~\ref{thm:radic-or-party}, every interior prior either is the
unique interior fixed point $p$, belongs to a separatrix, or
radicalizes. The singleton $\{p\}$ has measure zero, and by
Theorem~\ref{thm:party-measure-zero} so does the union of the separatrices.
Hence the set of priors which do not radicalize has measure zero.
\end{proof}

\begin{figure}[t]
    \centering
    \includegraphics[width=0.8\linewidth]{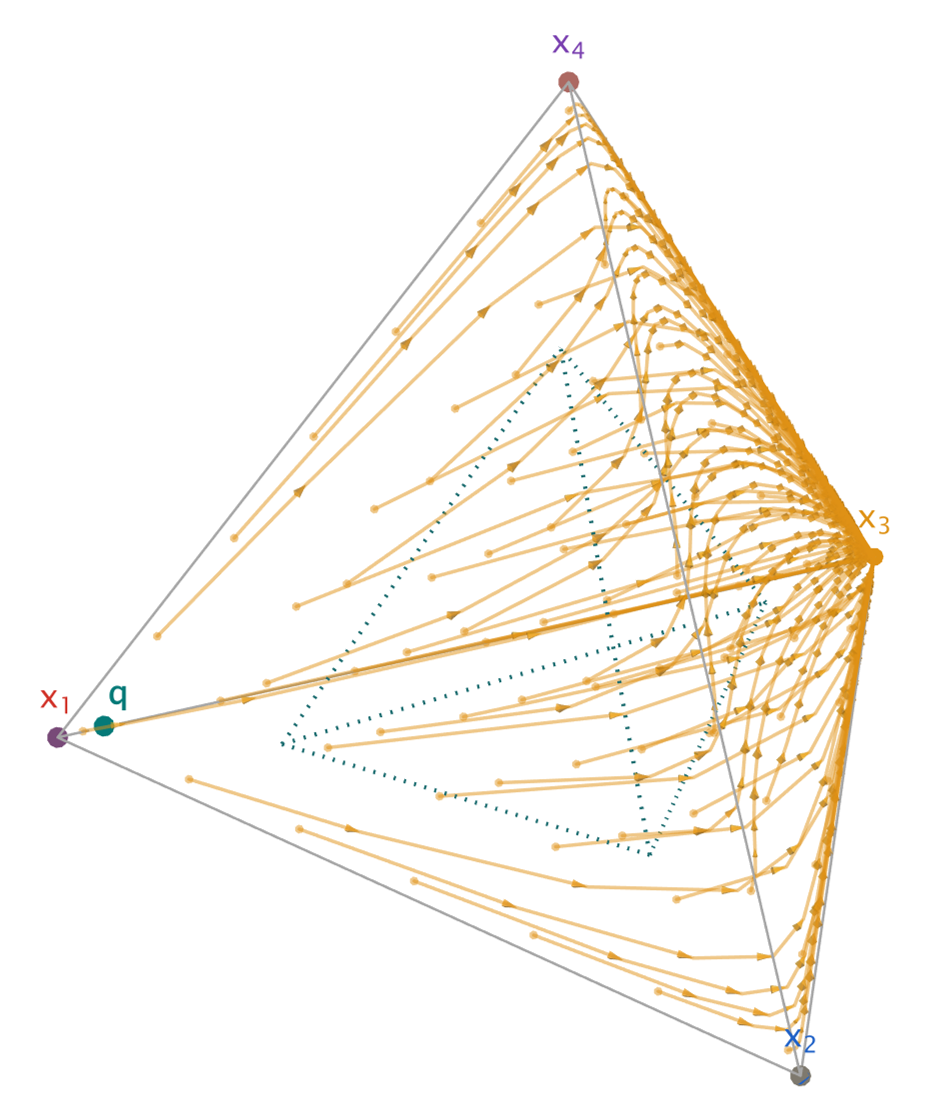}
    \caption{The iterated prior hacking trajectories for a quart channel with a choice of evidence $q$ that steers every interior prior to $\delta_{x_3}$. Theorem~\ref{thm:broadcast-common} gives a sufficient condition for this choice.}
    \label{fig:engine}
\end{figure}

\subsection{Can Shared Evidence Induce a Common Conclusion?}

With Theorem \ref{thm:party-measure-zero} and Corollary \ref{coro:almost-all-radicalize}, one may consider the range of possible Bayes phases in a simplex for a given channel (with the typical conditions of strict positivity and invertibility). Specifically, if one is able to tune $q$, which arrangements of Bayes phases are possible?

To interpret these questions analogically, consider an agent, Eve, who chooses the evidence $q$ received by the whole population. Now, clearly, for $q=\chn p$, agents who start in the interior and only repeatedly apply Jeffrey's update will all approach $p$. But what happens if she knows that the whole population will always do the opposite: that they \textit{always prior hack}? Suppose Eve can choose any full-support $q$. We say that she \emph{steers} a prior to a pure state if, with that choice of $q$ held fixed, iterated prior hacking takes the prior toward that state. Can she steer everyone to the same conclusion?

It turns out that she can choose a single $q$ for which \textit{every} interior prior radicalizes to the same pure state (Theorem~\ref{thm:broadcast-common}). That said, you cannot always prescribe \textit{which} pure state they end up in, as some may be precluded by the structure of the channel (Theorem \ref{thm:broadcast-not-any}). Nor can Eve always steer an individual to just \textit{any} pure state of her choice (Theorem \ref{thm:broadcast-individual}). We state these theorems formally.

\begin{theorem}
\label{thm:broadcast-common}
Let $\chn$ be strictly positive and invertible. Choose an
input $x_*$, by the following tie-breaking method. Look at the first row of $\chn$ and choose the input with the smallest entry. If several inputs tie, compare their entries in the second row and again choose the smallest. Continue through the rows, comparing only the inputs still tied. Since $\chn$ is invertible, no two columns are identical, so this eventually selects a single input. Now, with $x_*$ chosen, define
\[
b_x(y)=
\begin{cases}
\dfrac{\chn(y|x)-\chn(y|x_*)}
      {\max_{x'}\chn(y|x')},
& \chn(y|x)\ge\chn(y|x_*),\\[7pt]
\dfrac{\chn(y|x)-\chn(y|x_*)}
      {\min_{x'}\chn(y|x')},
& \chn(y|x)<\chn(y|x_*).
\end{cases}
\]
Any evidence $q$ satisfying the relations
\begin{align}
&\forall y: q(y)>0, \quad
\sum_y q(y)=1 \nonumber \\
&\forall(x\ne x_*): \sum_y q(y)b_x(y)>0 \label{eq:broadcast-sufficient} 
\end{align}
gives
\[
\mathcal{B}_{x_*}=\operatorname{int}\Delta.
\]
Such $q$ always exists for that $\chn$.
\end{theorem}
\begin{proof}
See Appendix~\ref{app:broadcast-common-proof} for the detailed proof and a worked example showing how to select $x_*$ and construct a single $q$ that steers every interior prior to $\delta_{x_*}$.
\end{proof}

In other words, Eve can \textit{always} find some range of $q$ such that every interior prior radicalizes to the same pure state, provided $\chn$ is invertible and strictly positive.

\begin{theorem}
\label{thm:broadcast-not-any}
There exist a strictly positive, invertible channel $\chn$ and an input $x$ such that, for every full-support evidence $q$, a nonempty open set of interior priors fails to radicalize to $\delta_x$.
\end{theorem}

\begin{proof}
See Appendix \ref{app:not-any-counter} for a counterexample establishing the result.
\end{proof}

\begin{theorem}
\label{thm:broadcast-individual}
There exist a strictly positive, invertible channel $\chn$ and an interior prior $\rf$ that cannot be radicalized to a prescribed pure state by any fixed full-support evidence $q$.
\end{theorem}

\begin{proof}
See Appendix \ref{app:not-any-counter} for a counterexample establishing the result.
\end{proof}

\section{Discussion}
\label{sec:discussion}
Now that we have studied the formal properties of these Bayesian trajectories under repeated evidence, we note some conceptual insights they offer. The analogical relevance of our illustrative setup may be situated in questions of belief formation and motivated reasoning in individuals \cite{kunda1990motivated,lord1979biased, Battaly2020-dogma-philo,cassam2016vice-dogma-philo,battaly2018dogmatism,battaly2018virtue,dunne2025vice,savion2009clinging-dogma-social,siebert2023effective-dogma-social,nguyen2020echo,taber2006motivated}. This can certainly be also extended to the context of \textit{collective} beliefs \cite{degroot1974consensus,kamenica2011persuasion}. 

With this in mind, we discuss four key points that emerge from this toy modeling of belief trajectories. 

\subsection{Repeated Prior Hacking and Radicalization}

In this work, we have understood closed-mindedness through the concretization of (i.e., ``doubling down'' on) beliefs. This was modeled by iterated prior hacking.  In each round of prior hacking, the prior that justified one conclusion becomes the conclusion to be justified next. Corollary~\ref{coro:almost-all-radicalize} tells us that almost every interior prior subjected to this procedure approaches a pure state. That is, the agent becomes certain of one possibility. This does not require the agent to begin particularly close to certainty: the result applies to almost every prior in the simplex interior. 

What happens to the fit with the evidence as the agent approaches certainty? Writing $\eta_i:=F_{\chn,q}^{-i}(\rf)$ for the prior-hacking trajectory, Theorem~\ref{thm:cpop-ineq} gives
\begin{equation}
    \mathcal{D}(\eta_{i+1})\geq\mathcal{D}(\eta_i),
    \label{eq:backward-KL}
\end{equation}
with strict inequality whenever $\eta_{i+1}\neq\eta_i$. That is, every nontrivial hacking step \textit{increases} $D_{\mathrm{KL}}(q\Vert\chn\rf)$. When $q=\chn p$, reversing the strict progress of Corollary~\ref{coro:cl-strict-progress} similarly increases $D_{\mathrm{KL}}(p\Vert\rf)$. Thus, although certainty is a statement about the limit, the worsening fit can be seen at every nontrivial step. That said, it is less obvious why this behavior has the particular structure defined by Theorem~\ref{thm:radic-or-party}, which we now discuss.

\subsection{What Separates the Bayes Phases?}

\subsubsection*{Facet Fixed Points as M-Projections}

Recall that the non-pure limiting states are the facet fixed points of Eq.~\eqref{eq:swing-mprojection}. Why should these arise from M-projections, minimizing $D_{\mathrm{KL}}(q\Vert\chn z)$, rather than I-projections, minimizing $D_{\mathrm{KL}}(\chn z\Vert q)$? This also concerns the direction of improvement under Bayesian updating in Theorem~\ref{thm:cpop-ineq}.

Now, there is certainly more than one way to explain this point, but we will focus on the lens of what distribution is taken as the \textit{population} and the \textit{sample} it generates. Note how Sanov's theorem \cite{sanov1957large} identifies $D_{\mathrm{KL}}(s\Vert r)$ as the large-deviation rate for empirical proportions near $s$ in an independent sample from population $r$. Smaller divergence means less exponential suppression as the sample grows.

For the M-projection, $q$ is the observed sample proportions, held fixed while we vary $z$, which is taken as the generating population (which, through $\chn$, gives the output population $\chn z$). Minimizing $D_{\mathrm{KL}}(q\Vert\chn z)$ essentially asks which candidate population $z$ makes the sample observations $q$ most plausible \footnote{For $n$ independent observations with empirical proportions $q$, the log likelihood satisfies $\log L(z)=n\large{\Sigma}_y q(y)\log[\chn z](y)=-n\mathcal{D}(z)+C$, where $C$ is independent of $z$. Thus the likelihood interpretation already holds at finite sample size.}. 

For the I-projection, the roles of $z$ and $q$ flip: $q$ is now taken as the hypothesized population, while $\chn z$ represents candidate empirical proportions. Minimizing $D_{\mathrm{KL}}(\chn z\Vert q)$ over $z\in S$ selects the least unlikely empirical distribution under the boundary conditions $\chn(S)$, in the large-deviation sense. 

Thus M-projection optimizes for a population $z$ (and therefore $\chn z$) that generates emergent observations represented by $q$, whereas I-projection would be optimizing for observations given by $\chn z$ compatible with a generating population $q$. The former matches our retrodictive setting: the evidence is supplied, and the prior (taken as the population that is generating the evidence) is revised to better account for it.

\subsubsection*{Bayes Phase Membership}

In connection to this, we can consider what the separation into Bayes phases tells us about these epistemic trajectories. Priors within the same phase do not need any particularly close, initial resemblance. Under repeated prior hacking, two vastly unlike beliefs can wind up radicalized to the same vertex (certainly, the most extreme case is implied by Theorem \ref{thm:broadcast-common}---where \textit{all} interior priors approach the same vertex via iterated prior hacking). 

Conversely, two initially similar beliefs on opposite sides of a separating separatrix may eventually become certain of entirely different alternatives (Sec.~\ref{sec:bayes-phases}, Figs.~\ref{fig:trit1} and~\ref{fig:quart1}). The Bayes phase membership therefore overrides any apparent doxastic similarity in such cases. As an analogy to tribalism or partisan division, this is quite specific. The agents can end up divided even though they receive the same evidence.

\subsubsection*{Agreement Without Crossing a Phase Boundary}

Now, it is certainly \textit{not} the case that prior hacking is the primary way to unite the descriptions of agents. The normative case of Jeffrey updating brings agents in different phases toward the same $p$. Interestingly, however, this is done without changing their phase membership (Theorem~\ref{thm:cl-emb-inv}, Sec.~\ref{sec:bayes-phases}). They can therefore approach agreement while retaining, at every finite step, different destinations should they return to an approach dominant in prior hacking. Agreement under rational updating is able to make irrelevant the distinctions that are consequential under prior hacking.

\subsubsection*{Separatrices in Higher Dimensions}

For trits, separatrices trace, in general, a curve (Fig.~\ref{fig:trit1} and \ref{fig:trit23}). For quarts, the separatrix (as defined in Eq.~\eqref{eq:party-line}) may form a surface (Fig.~\ref{fig:quart1}). This affords more ``wiggle room'' for arriving at a mixed state under iterated prior hacking. Properly, there are more ways for initial beliefs to differ while retaining the same eventual mixed conclusion. That said, there are also more alternatives for which agents could radicalize.

In this sense, introducing dimensions to the belief space gives rise to more possibilities for \textit{both} non-radicalizing (that is, mixed) \textit{and} radicalized beliefs. Nevertheless, the additional room to remain balanced certainly does not make this behavior significant. Even when separatrices form higher-dimensional surfaces, \textit{they remain thoroughly measure zero } with respect to the native simplex (Theorem~\ref{thm:party-measure-zero}): it is still the case that virtually all priors radicalize under repeated prior hacking (Corollary~\ref{coro:almost-all-radicalize}).

\subsection{How Much Does the Choice of Evidence Determine?}

Bayes phase structures being dependent not simply on $\chn$ but also on $q$ naturally raises the question of how much of it can be changed by choosing the evidence the population receives. For compatible evidence $q=\chn p$ with $p$ interior, all interior Jeffrey trajectories approach $p$ (Theorem~\ref{thm:cl-emb-inv}). More surprisingly, for prior hacking populations, this also occurs. Theorem~\ref{thm:broadcast-common} shows that for such $\chn$, there always exists some fixed $q$ that sends \textit{every} interior prior to the same pure state. However, Theorems~\ref{thm:broadcast-not-any} and~\ref{thm:broadcast-individual} show that one cannot always prescribe \textit{which} pure state this is. The latter is perhaps most unexpected: even if Eve only wishes to steer one specific individual to a chosen conclusion, there need not be \textit{any} fixed $q$ that accomplishes this \footnote{The obstruction is already visible in the divergence. In the counterexample of Appendix~\ref{app:not-any-counter}, any evidence satisfying a necessary condition for approach to the target places the initial prior at a larger divergence than the target itself. Eq.~\eqref{eq:backward-KL} then rules out reaching it.}. 

The distinction is between finding evidence that brings a population to agreement and finding evidence that brings it to a conclusion chosen in advance. The channel can rule out the latter even for a single individual.

\subsection{Neither Always Irrational nor Always Rational}

Finally, the two directions of updating need not describe two entirely separate kinds of agents. One might sometimes revise their beliefs through Jeffrey's update, and at other times prior hack. Since the two operations are inverses, their combined behavior is particularly simple under the same $\chn$ and $q$. Let $N_J(i)$ and $N_H(i)$ count the Jeffrey and hacking steps among the first $i$ updates. Writing $\zeta_i$ for the resulting mixed trajectory from $\rf_0$, we have
\begin{equation}
    \zeta_i=F_{\chn,q}^{\,N_J(i)-N_H(i)}(\rf_0).
    \label{eq:mixed-updates}
\end{equation}
This is nothing but the cancellation of one Jeffrey step by one hacking step, irrespective of their order.

Consequently, if $N_J(i)-N_H(i)\to+\infty$, the agent recovers $p$ when $q=\chn p$ (Theorem~\ref{thm:cl-emb-inv}); if it tends to $-\infty$, they instead approach the destination of uninterrupted prior hacking (Theorem~\ref{thm:radic-or-party}). Of course, whenever the numbers of the two updates are exactly equal, the agent returns to their initial prior.

All this is to say that an agent need not be consistently rational to recover the true state, nor consistently dogmatic to radicalize. Within this setting, a sufficient long-run tendency toward one operation still gives its corresponding destination. 

Now, as with our setup in general, keeping the same $q$ throughout is more of an analogical idealization, as is keeping the channel fixed. These assumptions allow the exact cancellation in Eq.~\eqref{eq:mixed-updates}; changing either generally removes it. What matters in this fixed setting is how the excess of Jeffrey steps over hacking steps (or vice versa) behaves as the number of updates grows.

\section{Conclusion}
\label{sec:conclusion}
Now that we have discussed the results in detail, we conclude. These emergent mathematical features give formal analogues to sociological phenomena familiar in the world \cite{nguyen2020echo,taber2006motivated}. Agents exposed to the same information can nevertheless approach certainty in different alternatives (Sec.~\ref{sec:bayes-phases}, also consider Figs.~\ref{fig:trit1},~\ref{fig:trit23} and~\ref{fig:quart1}). Repeated prior hacking generically leads toward maximal certainty despite a worsening fit to evidence (Corollary~\ref{coro:almost-all-radicalize}), while the non-pure alternatives require fine-tuning (Theorem~\ref{thm:party-measure-zero}). Such connections are particularly interesting under a ``Bayesian brains'' perspective, in which perception and cognition are understood through probabilistic inference and prediction \cite{knill2004bayesian,clark2013whatever,chivers2024-bayes-app}. This motivates asking how Bayesian descriptions might also illuminate problematic approaches to belief updating. Taken analogically, prior hacking \cite{aw-sidajaya-hacking} describes how a conclusion already held can be justified by changing the prior, connecting with questions of motivated reasoning and belief perseverance \cite{kunda1990motivated,lord1979biased}. The emergent information-theoretic features show how repeatedly engineering justification for preselected conclusions generically leads to certainty but at the cost of departing more and more from the evidence.

The sociological, psychological and even epistemological questions behind these behaviors are of interest in their own right \cite{savion2009clinging-dogma-social,taber2006motivated,Battaly2020-dogma-philo,cassam2016vice-dogma-philo}. Formal modeling within a Bayesian perspective gives an additional way to approach them, by making the assumptions and their consequences explicit. With toy models like this, one can formulate expectations about when initially similar beliefs diverge, when unlike beliefs eventually agree (Sec.~\ref{sec:bayes-phases} and Theorem~\ref{thm:cl-emb-inv}), and how far an agent can steer the population by determining the evidence it encounters (Theorems~\ref{thm:broadcast-common}--\ref{thm:broadcast-individual}). 

All this said, as mentioned in the introduction, our setting has essentially been the frictionless case so to speak: the evidence and channel remain static while only the prior changes. It would be worthwhile to consider varying $q$ and $\chn$, representing changing information and changing assumptions about how that information relates to the possibilities under consideration. The exact cancellation in Eq.~\eqref{eq:mixed-updates} would then generally be lost, and the order of rational updating and prior hacking could become consequential. To what extent radicalization persists, whether agents transition between the resulting evolving phases, and whether changing evidence can overcome the limits of steering with fixed evidence remain open questions. The present results provide a baseline for asking how problematic doxastic dynamics are structured and might be overcome.

\section*{Acknowledgments}
This project is supported by the National Research Foundation, Singapore through the National Quantum Office, hosted in A*STAR, under its Centre for Quantum Technologies Funding Initiative (S24Q2d0009); the Ministry of Education, Singapore, under the Tier 2 grant ``Bayesian approach to irreversibility'' (Grant No.~MOE-T2EP50123-0002), as well as the National Research Foundation, Singapore through the National Quantum Office, hosted in A*STAR, under its Quantum Engineering Programme 3.0 Funding Initiative (W24Q3D0001).

\section*{Disclosure}
During the preparation of this manuscript, the authors used generative Artificial Intelligence (AI) tools to assist with conceptual exploration, literature review, scientific analysis, coding, and drafting. All AI-assisted contributions were independently evaluated, verified and thoroughly edited by the authors, who take full responsibility for the scientific content and accuracy of the manuscript.


\bibliographystyle{unsrt}
\bibliography{apssamp}

\newpage
\appendix
\onecolumngrid
\section{Proofs}
\label{app:proofs}
\subsection{Proof for Theorem \ref{thm:cpop-ineq}} \label{app:cpop-ineq} 
 First, rewrite our conclusion:
\begin{eqnarray*}
   \eqref{eq:cpop-ineq} & \Leftrightarrow & \sum_{y} \cevi(y) (\log \cevi (y) - \log\,[\chn \rf_{i}](y))  \ge \, \, \sum_{y} \cevi(y) (\log \cevi (y) - \log\,[\chn\rf_{i+1}](y)) \\
    & \Leftrightarrow & \sum_{y} \cevi(y) \Big( \log \underbrace{[\chn\rf_{i+1}](y)}_{\sum_x \chn(y|x)\rf_{i+1}(x)} - \log\,[\chn\rf_i](y) \Big) \ge 0  \\ 
    & \Leftrightarrow & \sum_{y} \cevi(y) \Big( \log \sum_x \frac{\ret{\chn}{\rf_i}(x|y) }{\ret{\chn}{\rf_i}(x|y)}\chn(y|x) \rf_{i+1}(x) - \log\,[\chn\rf_i](y)\Big) \ge 0 
\end{eqnarray*}
Now, $\log \sum_k f(k) g(k) \geq \sum_k f(k) \log g(k)$ when $f(k)$ is always positive and $\sum_k f(k) =1$. Taking 
$\ret{\chn}{\rf_i}(x|y)$ in the numerator to play the role of $f(k)$, we have the implication that proving
\begin{eqnarray} 
    \sum_{y} & \cevi(y) \Big(   \sum_x \ret{\chn}{\rf_i}(x|y)  \log\frac{\chn(y|x) \rf_{i+1}(x)}{\ret{\chn}{\rf_i}(x|y)} \nonumber- \log[\chn\rf_i](y)\Big) \ge 0 \label{eq:int-pop-ineq}
\end{eqnarray}
implies the desired inequality \eqref{eq:cpop-ineq}. 
Notice now that by Bayes' rule, \eqref{eq:int-pop-ineq} is nothing but
\begin{align*}
        & \sum_{y} \cevi(y) \Big(  \sum_x \ret{\chn}{\rf_i}(x|y)  \log\frac{[\chn\rf_i](y) \, \rf_{i+1}(x)}{\rf_i(x)} - \log\,[\chn\rf_i](y)\Big) \ge 0 \\
        & \Leftrightarrow \; \sum_{y} \cevi(y)   \sum_x \ret{\chn}{\rf_i}(x|y) \log \frac{\rf_{i+1}(x)}{\rf_i (x)}   \ge 0 \\
        & \Leftrightarrow \; \sum_{x} \rf_{i+1}(x) \log \frac{\rf_{i+1}(x)}{\rf_i (x)} \ge 0 \Leftrightarrow \; {D_{\mathrm{KL}}}(\rf_{i+1}\Vert \rf_i) \ge 0
\end{align*}

According to Gibbs’ inequality, it is always true, and so \eqref{eq:cpop-ineq} obtains. The second equivalence comes from simply summing over $x$ for $\ret{\chn}{\rf_i}(x|y) \log [\chn\rf_i](y)$ which is just the second term in parentheses. 

Now, since ${D_{\mathrm{KL}}}(\rf_{i+1}\Vert \rf_i) = 0$ if and only if $\rf_{i+1}=\rf_i$, we also have the condition for the saturation:
\begin{equation}\label{eq:pop-ineq-strict}
    \rf_{i+1} \neq \rf_i \quad \Rightarrow \quad {D_{\mathrm{KL}}} \big(\cevi\Vert \chn \rf_{i+1}\big) < {D_{\mathrm{KL}}}\big(\cevi\Vert \chn \rf_{i}\big).
\end{equation}

\subsection{Proof for Theorem \ref{thm:radic-or-party}}
\label{app:radic-or-party}
This proof may be delineated in the following fashion. We first show that every accumulation point of the prior-hacking trajectory is fixed. Together with continuity and the finiteness of the fixed points, this implies that each trajectory goes to a unique fixed point. We then identify the possible limits: (1) an interior fixed point must be the true state, (2) a mixed limit on the boundary must be the facet fixed point of its supporting face, leaving (3) radicalization to a pure state. These give precisely the three alternatives in the theorem. Throughout, $\chn>0$ is an invertible stochastic matrix, $q>0$, and $\rf\in\operatorname{int}\Delta$. We use $\mathcal{D}$ as defined in Eq.~\eqref{eq:divergence}, and write
\begin{equation}\label{eq:convenience}
    \eta_i:=F^{-i}_{\chn,q}(\rf).
\end{equation}
Thus $\eta_0=\rf$ and $F_{\chn,q}(\eta_{i+1})=\eta_i$. With that, let's consider the steps in sequence.

\subsubsection*{Every Accumulation Point is a Fixed Point (that is Unique to its Subsequence)}
Now, an accumulation point $s$ of a trajectory, is any point for which that trajectory goes arbitrarily close to, at any arbitrary time (notice this is not the same as a \textit{fixed} point, as an accumulation point, for instance, could just alternate between itself and other points at some very large iteration length):
\begin{equation}
    \eta_{i_j}\longrightarrow s,
    \qquad i_j\longrightarrow\infty.
\end{equation}
So this does not yet require the whole trajectory to \textit{converge} to (i.e. stay at) some $s$. By Theorem~\ref{thm:cpop-ineq},
\begin{equation}
    \mathcal{D}(F_{\chn,q}(\rf))\leq \mathcal{D}(\rf),
\end{equation}
with equality if and only if $F_{\chn,q}(\rf)=\rf$. Consequently, the prior-hacking trajectory satisfies Eq.~\eqref{eq:backward-KL}, with strict inequality whenever $\eta_{i+1}\neq\eta_i$. Since $\chn(y|x)>0$, every component of $\chn\rf$ is bounded away from zero over the closed simplex. Hence $\mathcal{D}$ is continuous and bounded above. The sequence $\mathcal{D}(\eta_i)$ is therefore nondecreasing and convergent, so
\begin{equation}
    \mathcal{D}(\eta_{i+1})-\mathcal{D}(\eta_i)\longrightarrow0.
    \label{eq:KL-increments}
\end{equation}
Now, take any accumulation point $s$, with $\eta_{i_j}\to s$. Since $F_{\chn,q}(\eta_{i_j})=\eta_{i_j-1}$, we have
\begin{equation}
    \mathcal{D}(\eta_{i_j})-\mathcal{D}(F_{\chn,q}(\eta_{i_j}))
    =
    \mathcal{D}(\eta_{i_j})-\mathcal{D}(\eta_{i_j-1})
    \longrightarrow0.
\end{equation}
By continuity, this just means $\mathcal{D}(s)=\mathcal{D}(F_{\chn,q}(s)).$ The equality condition in Theorem~\ref{thm:cpop-ineq} then gives $F_{\chn,q}(s)=s$. Thus every accumulation point is fixed. The trajectory then converges to a \textit{single} fixed point, using the finiteness of the fixed points established below \footnote{Since every accumulation point is fixed, compactness and continuity imply that $\|\eta_{i+1}-\eta_i\|\to0$. Indeed, any convergent subsequence of $\eta_{i+1}$ approaches a fixed point $s$, and the corresponding preceding iterates $\eta_i=F_{\chn,q}(\eta_{i+1})$ approach the same $s$. Since there are finitely many fixed points, sufficiently late iterates lie in separated neighborhoods of these points. The vanishing step lengths prevent transitions between these neighborhoods. Thus the trajectory has a single accumulation point and, by compactness, converges to it.}.

\subsubsection*{The Only Three Possible Kinds of Fixed Points}
With that, we can now identify these fixed points. Every $s\in\Delta$ is either (1) in the interior, (2) in the relative interior of a positive-dimensional proper boundary face, or (3) a pure state. Let's consider these possibilities in turn. Recall that Jeffrey updating takes the component-wise form given in Eq.~\eqref{eq:jeffrey-component}. Let's apply it to our fixed points $s$. 
\begin{equation}
    [F_{\chn,q}(s)](x)
    =
    s(x)\sum_y
    \frac{\chn(y|x)q(y)}{[\chn s](y)}.
    \label{eq:jeffrey-component}
\end{equation}
Naturally, we have the following relation
\begin{equation}
    \sum_y
    \frac{\chn(y|x)q(y)}{[\chn s](y)}
    =1,
    \qquad x\in\operatorname{supp}s.
    \label{eq:fixed-factor}
\end{equation}

\subsubsection{Fixed Points in the Interior: They are always the true state $p$}
Suppose first that $s\in\operatorname{int}\Delta$. Eq.~\eqref{eq:fixed-factor} then holds for every $x$, so
\begin{equation}
    \chn^T\left(q\oslash[\chn s]\right)=\mathbf{1}.
\end{equation}
But $\chn^T\mathbf{1}=\mathbf{1}$ and $\chn$ is nonsingular.
Therefore $q\oslash[\chn s]=\mathbf{1}$. This is nothing but $q=\chn s$. Thus, if $q=\chn p$, injectivity gives $s=p$. Thus, there is at most one interior fixed point: the true state. This also implies that if $q$ fails to have a pre-image in $\operatorname{int}\Delta$, there is no interior fixed point (as we will see, this means prior hacking will always strictly push to the boundaries, even with immense fine-tuning). 

Furthermore, a trajectory can converge to $p$ only if its initial prior is already $p$ \footnote{Indeed, $\mathcal{D}(p)=0$, whereas $\mathcal{D}(\eta_i)$ is nonnegative and nondecreasing.}. This establishes the first alternative: all fixed points in the interior are the true state $p$ (if it exists). This trivial and very finely tuned scenario can be obtained if and only if one was prior hacking for something that actually was the truth all along.  

\subsubsection{Facet Fixed Points on the Supporting Faces}
Now suppose that $s$ is fixed and lies on the boundary with at least two nonzero coordinates. Let $S$ be the face whose relative interior contains $s$. Then $S$ is a positive-dimensional proper boundary face, and $s(x)>0$ for every state represented in $S$. We already know that $s$ is fixed. What remains is to show that it is the M-projection $s_S$. Differentiating $\mathcal{D}$ and using Eq.~\eqref{eq:fixed-factor}, we get
\begin{equation}
    \frac{\partial \mathcal{D}(s)}{\partial s(x)} =
    -\sum_y
    \frac{\chn(y|x)q(y)}{[\chn s](y)}
    =-1,
    \qquad x\in\operatorname{supp}s.
    \label{eq:KL-derivative}
\end{equation}
For any $z\in S$, both $z$ and $s$ are normalized and supported
on $S$. Consequently,
\begin{align}
    \sum_{x\in\operatorname{supp}s}
    \frac{\partial \mathcal{D}(s)}{\partial s(x)}
    \bigl(z(x)-s(x)\bigr)
    &=
    -\sum_{x\in\operatorname{supp}s}
    \bigl(z(x)-s(x)\bigr)
    =0.
\end{align}
Thus redistributing probability within $S$ produces zero
first-order change in $\mathcal{D}$. This is the constrained
stationarity condition on $S$. Since $q>0$ and $\chn$ is injective, $\mathcal{D}$ is strictly convex. Convexity therefore gives, for every $z\in S$,
\begin{align}
    \mathcal{D}(z) \geq \mathcal{D}(s)+ \sum_{x\in\operatorname{supp}s} \frac{\partial \mathcal{D}(s)}{\partial s(x)} \bigl(z(x)-s(x)\bigr) =\mathcal{D}(s).
\end{align}
Strict convexity makes this minimizer unique. This is nothing but the condition for a facet fixed point Eq.~\eqref{eq:swing-mprojection}. Namely,
\begin{equation}
    s
    =
    \arg\min_{z\in S}D_{\mathrm{KL}}(q\Vert\chn z)
    =
    s_S.
\end{equation}
In particular, there is at most one fixed point in the relative interior of each such face. If the trajectory converges to this fixed point, then $F^{-i}_{\chn,q}(\rf)\longrightarrow s_S,$ and therefore $\rf\in\mathcal{P}_S$ by
Eq.~\eqref{eq:party-line}. This establishes the second alternative: all non-pure fixed states not in the interior are facet fixed points of the boundary in which they reside.

\subsubsection{Pure States as Fixed Points}
This leaves pure states, i.e. $s=\delta_x$ for some $x$. Then $ F^{-i}_{\chn,q}(\rf)\longrightarrow\delta_x,$ which is precisely radicalization as defined in Eq.~\eqref{eq:rad}. This establishes the third alternative.

\subsubsection*{Conclusion}
The classification gives at most one interior fixed point,
at most one fixed point in the relative interior of each
positive-dimensional proper boundary face, and finitely
many vertices. Thus there are finitely many fixed points,
as required by the convergence argument above. Together, every interior prior converges to a unique fixed point. If that limit is interior, the initial prior was already the true state. If it is non-pure and on the
boundary, the initial prior belongs to a separatrix leading to a facet fixed point. If it is pure, the initial prior radicalizes. These possibilities are mutually exclusive and exhaust the simplex interior, so exactly one of the three alternatives in the theorem obtains.

\hfill$\square$

\subsection{Proof for Theorem \ref{thm:party-measure-zero}} \label{proof:party-measure-zero}
Appendix~\ref{app:radic-or-party} already establishes convergence and identifies the possible limits. The question here is how finely tuned an interior prior must be to converge to a (non-pure) facet fixed point. We first show that such convergence imposes lower-dimensional restrictions (i.e. the source of the fine-tuning) near the limiting state. We then trace these restrictions back to the initial prior. Together, this level of fine-tuning gives rise to the measure-theoretic negligibility of the separatrices.

\subsubsection*{Priors on a Separatrix Approach Its Facet Fixed Point}
By definition \eqref{eq:party-line}, any nonempty separatrix $\mathcal{P}_S$ associated with a non-pure facet fixed point $s_S$ is such that, for any $\rf\in\mathcal{P}_S$, $\eta_i:=F^{-i}_{\chn,q}(\rf)\longrightarrow s_S.$ Every finite iterate remains in the interior of the simplex $\Delta$, even though its limit lies on the boundary. Since the trajectory converges, all sufficiently late iterates must remain near $s_S$. With this, the question is then which nearby interior priors can stay there. 

\subsubsection*{Remaining near implies fine-tuning}
To see which perturbations contract under Jeffrey updating, we first restrict the derivative to the supporting face.  By Appendix~\ref{app:radic-or-party}, $s_S$ is fixed and satisfies Eq.~\eqref{eq:fixed-factor}. Differentiating the Jeffrey update at $s_S$, for $x,x'\in\operatorname{supp}(s_S)$ we obtain
\begin{equation}
    \left.\frac{\partial [F_{\chn,q}(\rf)](x)}{\partial\rf(x')}\right|_{\rf=s_S}
    =
    \delta_{xx'}
    -
    s_S(x)\sum_y
    \frac{q(y)\chn(y|x)\chn(y|x')}{[\chn s_S](y)^2}.
    \label{eq:party-derivative}
\end{equation}

Consider a nonzero perturbation $v$ supported on $\operatorname{supp}(s_S)$ with $\sum_xv(x)=0$. Injectivity, followed by Cauchy--Schwarz and Eq.~\eqref{eq:fixed-factor}, gives
\begin{equation}
    0
    <
    \sum_y q(y)\left(\frac{[\chn v](y)}{[\chn s_S](y)}\right)^2
    <
    \sum_{x\in\operatorname{supp}(s_S)}
    \frac{v(x)^2}{s_S(x)}.
\end{equation}
The upper inequality is strict because $\chn>0$: equality would require $v(x)/s_S(x)$ to be constant on the support, which together with $\sum_xv(x)=0$ would give $v=0$. In the rescaled coordinates $v(x)/\sqrt{s_S(x)}$, the derivative in Eq.~\eqref{eq:party-derivative} is symmetric. These bounds therefore place its eigenvalues along the supporting face \textit{strictly} between $0$ and $1$.

Thus Jeffrey updating contracts the $|\operatorname{supp}(s_S)|-1$ independent directions tangent to the supporting face at $s_S$. The subtraction of one accounts for normalization.
Now, reversing this \footnote{We confirm local invertibility. For $x\notin\operatorname{supp}(s_S)$, the corresponding derivative row has only one nonzero entry,
$
    \left.\frac{\partial [F_{\chn,q}(\rf)](x)}{\partial\rf(x)}\right|_{\rf=s_S}
    =
    \Sigma_y  \frac{\chn(y|x)q(y)}{[\chn s_S](y)}
    >0.
$
Together with the nonzero eigenvalues along the face, this makes the derivative nonsingular on the affine hyperplane containing $\Delta$. The update formula extends smoothly to a neighborhood of $s_S$ in this hyperplane, since its denominators remain positive. Hence the inverse function theorem gives a smooth local inverse.} we obtain the behavior of prior hacking at such perturbations. The upshot is that prior hacking would \textit{expand} perturbations that Jeffrey updating contracts. 
In particular, there are at least $|\operatorname{supp}(s_S)|-1$ expanding directions at $s_S$. This is the role of the supporting face: it identifies directions in which nearby trajectories can depart from the proposed limit. In these directions the inverse map expands perturbations to first order.

\subsubsection*{The permitted interior priors occupy a lower-dimensional set}
The local center-stable manifold theorem \cite{lee2016gradient} makes this restriction precise; see also \cite{wen2016differentiable, smale1967differentiable} for background on stable manifolds. It gives a local manifold $\mathcal{C}_S$ containing every state whose subsequent hacking iterates remain sufficiently close to $s_S$, with
\begin{equation}
    \dim\mathcal{C}_S
    \leq
    (d-1)-\bigl(|\operatorname{supp}(s_S)|-1\bigr)
    =
    d-|\operatorname{supp}(s_S)|.
    \label{eq:party-dimension}
\end{equation}
Since $s_S$ is non-pure, $|\operatorname{supp}(s_S)|\geq2$, and therefore
\begin{equation}
    \dim\mathcal{C}_S\leq d-2<d-1.
\end{equation}
This is the precise fine-tuning condition. The nearby interior priors that can continue approaching $s_S$ must satisfy at least one independent local constraint. They need not lie on the boundary face: $\mathcal{C}_S$ may extend into the interior. But its dimension is smaller than that of the interior, so
\begin{equation}
    \mu\bigl(\mathcal{C}_S\cap\operatorname{int}\Delta\bigr)=0.
\end{equation}

\subsubsection*{Concluding the Proof}
Every prior on $\mathcal{P}_S$ eventually enters the local measure-zero set $\mathcal{C}_S\cap\operatorname{int}\Delta$. It can therefore be recovered from that set by finitely many Jeffrey updates. Each finite iterate preserves measure zero, since its derivative is bounded on the compact simplex. Taking the countable union over all possible numbers of steps gives $\mu(\mathcal{P}_S)=0$. Finally, Appendix~\ref{app:radic-or-party} gives only finitely many distinct mixed boundary limits. Hence
\begin{equation}
    \mu\left(\bigcup_S\mathcal{P}_S\right)=0.
\end{equation}
Together, convergence to a non-pure facet fixed point requires the interior trajectory to satisfy lower-dimensional constraints near its limit. Tracing these constraints back to the initial prior preserves measure zero. Thus separatrices are possible, but membership in one requires precisely this fine-tuning.

\hfill$\square$

\subsection{Proof for Theorem \ref{thm:broadcast-common}} \label{app:broadcast-common-proof}

Each $[\chn\rf](y)$ lies between the smallest and largest
entries in row $y$. Consequently,
\[
\sum_y
\frac{q(y)[\chn(y|x)-\chn(y|x_*)]}{[\chn\rf](y)}
\ge
\sum_y q(y)b_x(y)
=:c_x>0
\]
uniformly over $\rf\in\Delta$.

By Eq.~\eqref{eq:jeffrey-component}, the inverse iterates
$\eta_i=F_{\chn,q}^{-i}(\rf)$ therefore satisfy
\[
\frac{\eta_{i+1}(x)}{\eta_{i+1}(x_*)}
=
\frac{\eta_i(x)}{\eta_i(x_*)}
\frac{\displaystyle\sum_y
 q(y)\chn(y|x_*)/[\chn\eta_{i+1}](y)}
{\displaystyle\sum_y
 q(y)\chn(y|x)/[\chn\eta_{i+1}](y)}.
\]
The denominator in the last factor exceeds its numerator
by at least $c_x$. Strict positivity of $\chn$ bounds
the numerator uniformly above, so this factor is uniformly
less than one. Repeated inverse updates thus send
$\eta_i(x)/\eta_i(x_*)$ to zero for every $x\ne x_*$.
Hence every interior prior converges to $\delta_{x_*}$.

Now, for completeness, we show that an appropriate $q$ adhering to Eq.~\eqref{eq:broadcast-sufficient} always exists. For each $x\ne x_*$, the first nonzero value of $b_x(y)$ is positive. Choose positive weights successively, making each new weight small enough that later contributions do not overturn any positive sum already obtained. This is possible because there are only finitely many inequalities to preserve. 

Every competing column eventually differs from that of $x_*$, so all sums $\sum_y w(y)b_x(y)$ are ultimately positive. Normalizing the weights gives the required full-support evidence $q$.

\subsubsection*{Example of the Tie-Breaking Construction}
Consider the channel
\[
\chn=\frac1{10}
\begin{pmatrix}
1&1&1&2\\
2&2&3&1\\
2&3&1&2\\
5&4&5&5
\end{pmatrix},
\]
with columns indexed by $x_1,\ldots,x_4$ and rows by
$y_1,\ldots,y_4$. Its entries are strictly positive, its columns
sum to one, and $\det\chn=1/1000$.

The first row leaves $x_1,x_2,x_3$ tied for the smallest entry,
excluding $x_4$. Comparing only these three inputs in the second
row leaves $x_1,x_2$. The third row then selects $x_1$, so
$x_*=x_1$. Notice that $x_4$ has the smallest entry in the second
row, but it has already been excluded. Likewise, $x_3$ has the
smallest entry in the third row, but is no longer a candidate.
Thus the procedure does not require one column to minimize every row.

With this choice of $x_*$, the coefficients in
Theorem~\ref{thm:broadcast-common} are
\[
\begin{array}{c|rrrr}
 & y_1 & y_2 & y_3 & y_4\\ \hline
b_{x_2}(y)&0&0&\frac13&-\frac14\\[3pt]
b_{x_3}(y)&0&\frac13&-1&0\\[3pt]
b_{x_4}(y)&\frac12&-1&0&0
\end{array}.
\]
For example, the negative entry $b_{x_2}(y_4)$ uses the
smallest entry of the fourth row, giving
$(4/10-5/10)/(4/10)=-1/4$.
Writing $w_j=w(y_j)>0$, the sufficient inequalities become
\[
\frac{w_3}{3}-\frac{w_4}{4}>0,\qquad
\frac{w_2}{3}-w_3>0,\qquad
\frac{w_1}{2}-w_2>0.
\]
These make the successive choice of weights explicit. Start with
$w_1=24$. Choose $w_2=8< w_1/2$, then
$w_3=2<w_2/3$, and finally $w_4=1<4w_3/3$.
Each choice preserves the inequalities already made positive.
Normalizing gives the full-support evidence
\[
q=\frac1{35}(24,8,2,1)^{\mathsf T}.
\]
Indeed,
\begin{align*}
\sum_y q(y)b_{x_2}(y)
&=\frac1{35}\left(\frac23-\frac14\right)=\frac1{84}>0,\\
\sum_y q(y)b_{x_3}(y)
&=\frac1{35}\left(\frac83-2\right)=\frac2{105}>0,\\
\sum_y q(y)b_{x_4}(y)
&=\frac1{35}\left(12-8\right)=\frac4{35}>0.
\end{align*}
Thus Eq.~\eqref{eq:broadcast-sufficient} holds, and
Theorem~\ref{thm:broadcast-common} gives
\[
F_{\chn,q}^{-i}(\rf)\longrightarrow\delta_{x_1}
\qquad\text{for every }\rf\in\operatorname{int}\Delta.
\]
For this channel, Eve can therefore hold this single $q$ fixed
and steer every interior prior to the same pure state.

\subsection{Proof for Theorem \ref{thm:broadcast-not-any} and Theorem \ref{thm:broadcast-individual}} \label{app:not-any-counter}
Consider
\[
\chn=\frac1{10}
\begin{pmatrix}
8&1&4\\
1&8&4\\
1&1&2
\end{pmatrix},
\]
which is strictly positive and has determinant $7/100$. We show that no choice of $q$ makes every interior prior radicalize to $\delta_3$. For each output $y$, the third column satisfies $\chn(y|3)>\sqrt{\chn(y|1)\chn(y|2)}$. With $\mathcal{D}$ as defined in Eq.~\eqref{eq:divergence}, we have
$\mathcal{D}(\delta_x)=\sum_y q(y)\log\frac{q(y)}{\chn(y|x)}$.
One can then verify that taking logarithms before averaging with $q$ gives
\[
\mathcal{D}(\delta_3)
<
\frac{\mathcal{D}(\delta_1)+\mathcal{D}(\delta_2)}2.
\]

Thus at least one of $\delta_1,\delta_2$ has larger divergence than $\delta_3$. By continuity, a nonempty open set of interior priors near that vertex also satisfies $\mathcal{D}(\rf)>\mathcal{D}(\delta_3)$. Since prior hacking cannot decrease $\mathcal{D}$, none of these priors can go to $\delta_3$. This concludes the proof for Theorem \ref{thm:broadcast-not-any}.

With the same $\chn$, take $\rf=(9/10,1/20,1/20)$ with target $\delta_3$. We show that any evidence permitting convergence near $\delta_3$ makes the initial divergence too large to reach it.

Suppose $\eta_i=F_{\chn,q}^{-i}(\rf)\to\delta_3$. By Eq.~\eqref{eq:jeffrey-component},
\[
\frac{\eta_{i+1}(2)}{\eta_i(2)}
\longrightarrow
\left(\frac14q(1)+2q(2)+\frac12q(3)\right)^{-1}.
\]
This limit cannot exceed one, since otherwise the positive coordinate $\eta_i(2)$ would eventually increase rather than tend to zero. Using $\sum_y q(y)=1$, convergence therefore requires $q(2)\ge\frac{3-q(3)}7.$ However, $\chn\rf=(149,30,21)/200$, so
\[
\mathcal{D}(\rf)-\mathcal{D}(\delta_3)
=
q(1)\log\frac{80}{149}
+q(2)\log\frac83
+q(3)\log\frac{40}{21}.
\]
After substituting $q(1)=1-q(2)-q(3)$, this expression increases with $q(2)$. Substituting its lower bound above gives an expression that also increases with $q(3)$. Hence
\[
\mathcal{D}(\rf)-\mathcal{D}(\delta_3)
\ge
\frac47\log\frac{80}{149}
+\frac37\log\frac83
>0.
\]
Thus the initial divergence already exceeds that of the target. Since prior hacking cannot decrease $\mathcal{D}$, convergence to $\delta_3$ is impossible. This gives the counterexample required for Theorem \ref{thm:broadcast-individual}.

\end{document}